\documentclass[pra, twocolumn, showpacs, superscriptaddress]{revtex4-1}
\usepackage{graphicx,amsmath}
\usepackage{color}
\usepackage{stackengine}

\begin{document}

\title{Collective Dynamics and Atom Loss in Bright Soliton Matter Waves}
\author{Daniel Longenecker}
\affiliation{Laboratory of Atomic and Solid State Physics, Cornell University, Ithaca, New York 14853, USA}
\author{Erich J. Mueller}\email{em256@cornell.edu}
\affiliation{Laboratory of Atomic and Solid State Physics, Cornell University, Ithaca, New York 14853, USA}
\date{\today}

\begin{abstract}


Motivated by recent experiments, we model the dynamics of bright solitons formed by cold gases in quasi-1D traps.  A dynamical variational ansatz captures the far-from equilibrium excitations of these solitons. Due to a separation of scales, the radial and axial modes decouple, allowing for closed-form approximations for the dynamics. We explore how soliton dynamics influence atom loss, and find that the time-averaged loss is largely insensitive to the degree of excitation. The variational approach enables us to perform high precision calculations of the critical atom number (ie. the maximum number of atoms that can exist in a single soliton before the attractive forces overwhelm quantum pressure, leading to collapse).

\end{abstract}
\maketitle
\section{Introduction}

Solitons are localized excitations of nonlinear media which robustly maintain their shape \cite{kasman2010,drazin1989}.  Due to their robustness they play key roles in settings as disparate as weather patterns, optical fibers, and quantum degenerate atoms \cite{remoissenet1999}. The latter has provided a particularly large degree of control, enabling the quantitative investigation of a wide range of soliton phenomena \cite{abdullaev2008}. For example, cold atom solitons can collapse and explode in a ``Bosenova'' \cite{cornish2006}, form mesoscopic Bell states \cite{gertjerenken2013}, and improve the noise floor of a matter-wave interferometer \cite{polo2013}. Here we model the far from equilibrium dynamics of a single bright soliton formed from an atomic gas in a highly anisotropic trap. 

This study is motivated by experiments at Rice University where a quantum degenerate gas of bosonic lithium atoms is confined by a quasi-1D trap \cite{hulet2017}. The interactions are suddenly changed from repulsive to attractive, which makes the cloud unstable. This modulational instability causes the cloud to break up into a train of solitons. The solitons bounce off one another and undergo large shape deformations. Here we model the dynamics of the shape deformations of a single soliton in this train. We use a time dependent variational method to derive equations of motion for parameters which describe the soliton's shape. We furthermore make a series of approximations which yield closed-form expressions for the length and thickness of the soliton. For much of the experimentally relevant parameter range, we find that the closed-form expressions provide a good approximation to the dynamics. 

Due to three-body processes, atoms are continually scattered out of the solitons and lost from the trap. Interestingly, the experimental loss rate appears to be orders-of-magnitude faster than would be expected from models of static solitons. One possible hypothesis is that the extra loss is due to the large-scale oscillations of the solitons. We model this process, exploring how atom loss is influenced by the dynamics. In the regime of large oscillations, collisions predominantly occur when the cloud is most compressed. We quantify these episodic losses and characterize their dependence on the amplitude of the oscillations. Radial oscillations slightly increase the loss rate, while axial oscillations slightly decrease it. Since the size of the effect is relatively small, we conclude that this hypothesis is not able to account for the greatly enhanced loss observed in the experiment.

The nonlinear dynamics literature contains some discussion of dynamical solitons -- particularly ``breathers," which are a class of time dependent solutions to equations such as the nonlinear Schrodinger equation. Breathers are important for understanding phenomena such as rogue waves, and have some connections to the physics explored here. 

\begin{figure}
\includegraphics[width=\columnwidth]{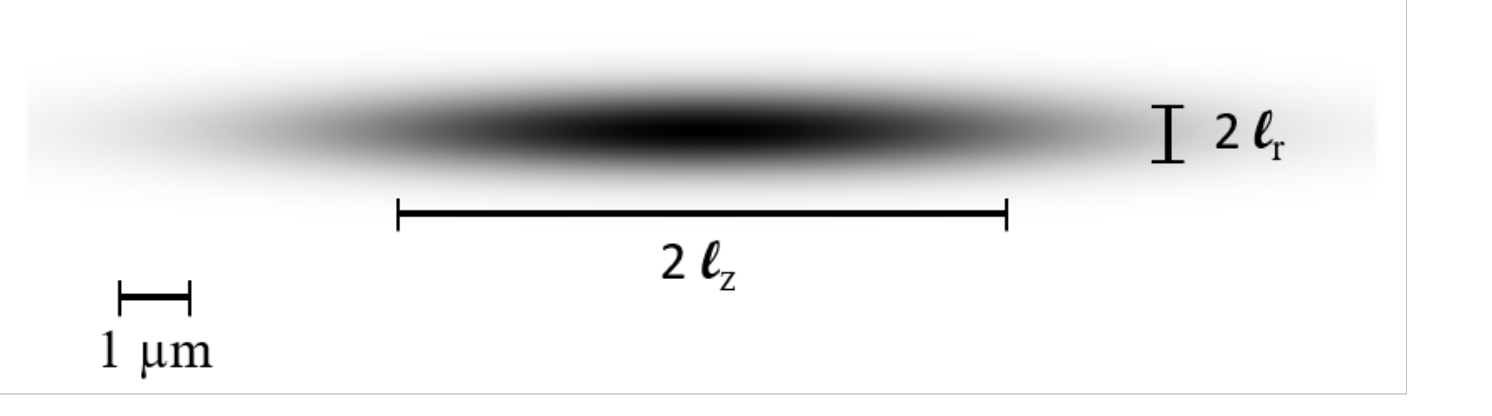}
\caption{\label{atomcloud} Geometry of soliton-shaped atomic cloud. Scale bar shows typical size in experiments \cite{hulet2017}.}\label{soltioncloud}
\end{figure}

We use two distinct approaches to modeling the soliton dynamics. The first, and most conventional, is to discretize space and numerically integrate the Gross-Pitaevskii equation. The second is to write down an ansatz wavefunction and derive a set of coupled ordinary differential equations for the parameters from a time-dependent variational principle. The second approach is much more numerically efficient than the first, as it identifies the most important degrees of freedom. We systematically increase the number of parameters in the ansatz. The six-parameter Gaussian ansatz captures the qualitative features, while the sixteen-parameter ansatz is more accurate than our spatial discretization approach (using a grid-spacing appropriate for calculations on a laptop computer).  Using a separation-of-scales, we simplify the equations for the Gaussian ansatz, and make a formal mapping onto two classic problems, simple harmonic and planetary motion.  We thereby produce approximate closed-form solutions to the equations and find invariants 
 which have physical significance.

Throughout this paper we model a single soliton. By contrast, the most recent experiments at Rice University study a train of 10-20 solitons. Although generalizing our results to a soliton train is straightforward, it is beyond the scope of this study. Qualitative features, such as the dependence of loss-rate on excitation amplitude, are nonetheless directly relevant to these experiments \cite{hulet2017}.

\begin{figure}
\includegraphics[width=\columnwidth]{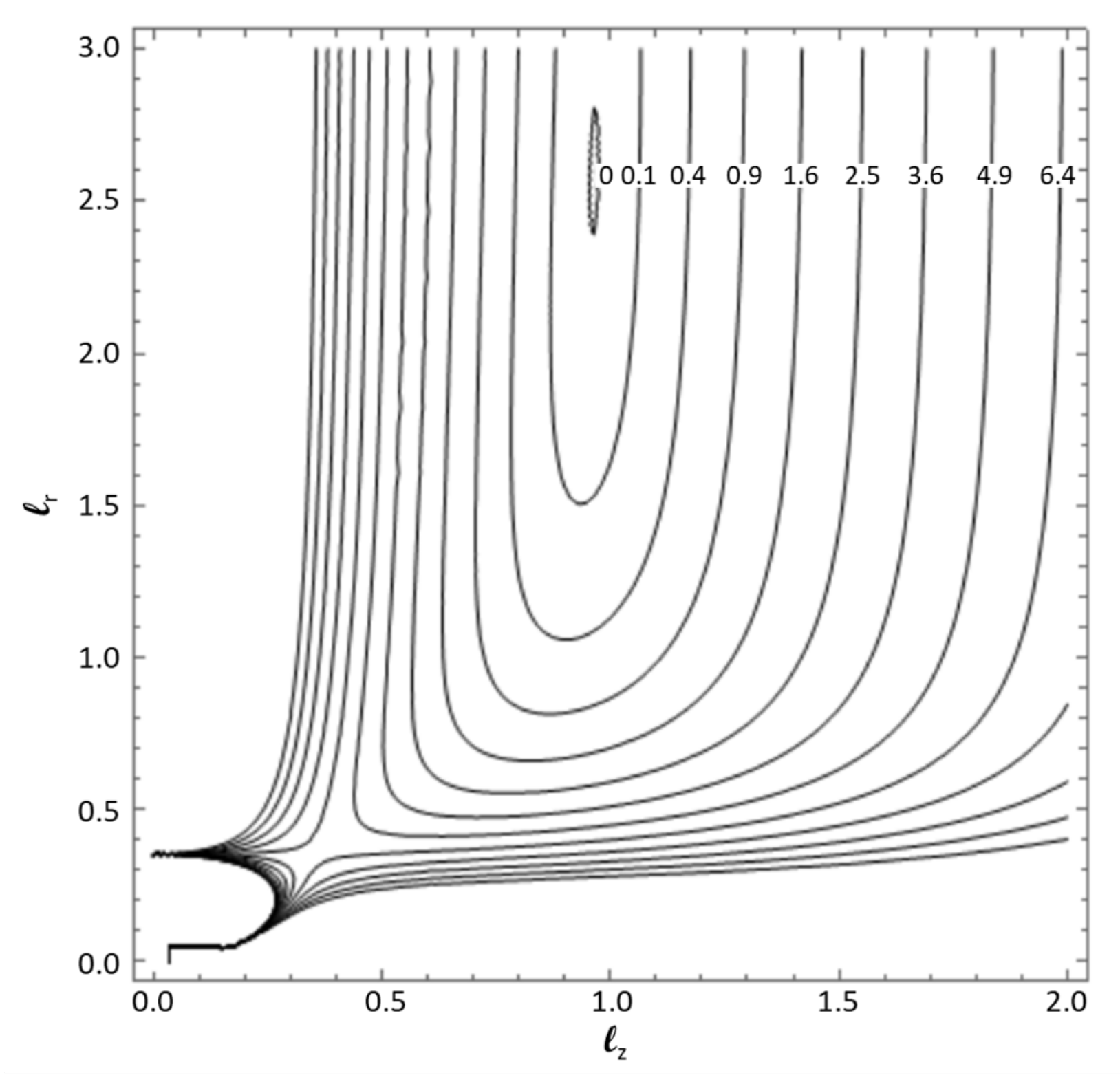}
\caption{\label{solitoncontours} Energy landscape of a soliton within the Gaussian approximation. Length-scales and energies are measured in transverse oscillator units. Parameters: $\omega_r/\omega_z=50$, and $N/N_c=0.64$. Contours are labeled by energies, and spaced quadratically.}\label{efig}
\end{figure}

\section{Model}

The cold gas experiments of \cite{hulet2017} are well-modeled by 
the three-dimensional, axially symmetric Gross-Pitaevskii equation with a harmonic trap and three-body loss \cite{pethick2008},
\begin{multline}\label{gpe1}
i \hbar \partial_{\tilde t} \tilde\psi = -\frac{\hbar^2}{2m}{\tilde \nabla}^2 \tilde\psi + \frac{m}{2}({\tilde\omega_r}^2 \tilde{r}^2+{\tilde\omega_z}^2 \tilde{z}^2) \tilde{\psi}\\
+ \tilde g |\tilde \psi|^2\tilde \psi - i \tilde\Gamma_3 |\tilde\psi|^4\tilde\psi,\end{multline}
where the tilde symbols are used to denote quantities with physical units. Planck's constant is $\hbar$, and the atomic mass is $m$. The axial trap frequency $\tilde\omega_z$ is much smaller than the radial frequency $\tilde\omega_r$: in the experiments of \cite{hulet2017}, $\frac{\tilde\omega_r}{2\pi}=346$ Hz, and $\frac{\tilde\omega_z}{2\pi}=7.4$ Hz. For the soliton dynamics that we study, the axial trap plays a very small role. Thus, we set $\tilde \omega_z=0$ for most of the discussion. The atoms interact via a short-range potential with scattering length $a_s$, yielding a coupling constant $\tilde g=4\pi\hbar^2 a_s/m$. In the experiments, $a_s$ is tuned via a Feshbach resonance \cite{chin2010} from approximately $3a_0$ to various negative values of order $-a_0$, where $a_0$ is the Bohr radius. Atom losses from inelastic three-body collisions are modeled by the term proportional to $\Gamma_3$, where $\Gamma_3=L_3/12\sim10^{-29}$ cm$^6$/s. Physically, the atomic density is $n(r)=|\psi(r)|^2$. We adimensionalize this equation, scaling length as $x=\sqrt{\frac{m \tilde\omega_r}{\hbar}}\tilde x,$ time as $t=\tilde\omega_r\tilde t,$ frequency as $\omega=\frac{\tilde\omega}{\tilde\omega_r},$ and the wavefunction as $\psi=(\frac{\hbar}{m \tilde\omega_r})^\frac{3}{2}\tilde\psi$. The dimensionless coupling constants are $g=4 \sqrt{2\pi^3}\hbar\tilde\omega_r (\frac{\hbar}{m \tilde\omega_r})^\frac{3}{2} \tilde g$ and $\Gamma_3=m^3 \tilde\omega_r^2 \hbar^{-3} \tilde\Gamma_3$

We study systems with attractive interactions: $a_s<0$. In the absence of loss, Eq.~(\ref{gpe1}) will have a metastable soliton solution as long as $N<N_c=k_c\sqrt{\frac{\hbar}{m \tilde\omega_r}}\frac{1}{|a_s|}$ \cite{perezgarcia1998,parker2007}. In Sec.~\ref{sec:numerical}, we calculate that $k_c=0.677986(2)$ when $\omega_z=0$. As explained in \cite{billam2012}, this equilibrium physics can be understood through a Gaussian ansatz $\psi\propto \exp(-r^2/(2 \ell_r^2)-z^2/(2\ell_z^2))$. The energy as a function of $\ell_r$ and $\ell_z$ is 

\begin{equation}
E_{\rm gauss}=\frac{N}{2}\left(\frac{1}{\ell_r^{2}}+\frac{1}{2\ell_z^2}+\ell_r^2+ \frac{gN}{\ell_r^2\ell_z}\right)
\end{equation}

An example energy landscape is shown in Fig.~\ref{efig} for a particular $N<N_c$.  It consists of two basins, separated by a saddle. In the region where both $\ell_z$ and $\ell_r$ are small, the attractive interactions dominate, causing the cloud to shrink rapidly. Consequently, the density becomes very high and rapid three-body recombination leads to large atom losses \cite{wieman2001,hulet2013}. This process of shrinking and evaporating is often referred to as a collapse. When $N<N_c$ the energy landscape also contains a local minimum at finite $\ell_z$ and $\ell_r$ where the soliton is metastable. The energy landscape is highly anisotropic around this minimum: the curvature of the energy landscape in the $\ell_z$ direction is much smaller than the curvature in the $\ell_r$ direction. In the limit $N\ll N_c$, the low energy dynamics become purely axial, and Eq.~(\ref{gpe1}) reduces to a 1D nonlinear Schrodinger equation \cite{billam2012}. 

When $N>N_c$  the energy landscape is qualitatively different, with no stationary points at nonzero $\ell_z$ and $\ell_r$. In this limit a single soliton will always collapse.  As is illustrated by the experiment, the only way to have $N>N_c$ is to have multiple solitons.

To model the dynamics of a soliton, we extend the ansatz to a completely general time-dependent form,
\begin{multline}\label{ansatz1}
\psi(r,z,t) = \frac{A(r,z,t)}{\pi^{\frac{3}{4}} \ell_z(t)^{\frac{1}{2}} \ell_r(t)} \exp[-\frac{r^2}{2\ell_r(t)^2}-\frac{z^2}{\ell_z(t)^2}]\\\exp[ir^2\phi_r+ iz^2 \phi_z],
\end{multline}
where again $\ell_r(t)$ is the radial size, and $\ell_z(t)$ is the axial size. The phases proportional to $r^2$ and $z^2$ are related to currents. 

The factor $A(r,z,t)$ determines the number of particles in the soliton and accounts for all deviations from the Gaussian shape.  We impose cylindrical and inversion symmetry.  Consequently, we expand $A$ in a set of even-degree orthogonal polynomials, up to total degree $2n$:
\begin{eqnarray}\nonumber
A
&=&\displaystyle\sum_{l=0}^n\sum_{m=-\frac{l}{2}}^{\frac{l}{2}} c_{lm}(t)H_{l-2m}\left(\frac{z}{\ell_z}\right)L_{\frac{l}{2}+m}\left(\frac{r^2}{\ell_r^2}\right)\gamma_{lm}\\\label{number1}
&&\gamma_{lm}=\sqrt{\frac{2^{m-\frac{l}{2}}}{(\frac{l}{2}-m)!}},\end{eqnarray} where $H_{k}(z)$ is the $k^{th}$ Hermite polynomial and $L_k(r)$ is the $k^{th}$ Laguerre polynomial. The factors are chosen so that \begin{equation}N=\int \!d^3x\, |\psi|^2 =\sum_{l=0}^n\sum_{m=-\frac{l}{2}}^{\frac{l}{2}} |c_{lm}|^2.
\end{equation}
In the limit $n\to\infty$, the exact solution to Eq.~(\ref{gpe1}) can be written in the form of the ansatz in Eq.~(\ref{number1}).  By increasing $n$, we can increase the accuracy of our approximation, at the expense of increasing the number of coupled equations which need to be solved.

To derive the equations of motion for our variational parameters we write
the Gross-Pitaevskii equation, Eq.~(\ref{gpe1}), as 
\begin{equation}\label{vareom}\frac{\delta S}{\delta \psi^*}=-i \Gamma_3 |\psi|^4\psi\end{equation}
where the action $S$ is a functional of $\psi$,
\begin{eqnarray}\label{s}
S&=&\int dt~ d^3x ~
\left[ i \psi^* \partial_t \psi -{\cal H}\right]\\
{\cal H} &=&
\frac{1}{2}|\nabla\psi|^2 +\frac{1}{2}(r^2+\omega_z^2 z^2) |\psi|^2 -\frac{1}{2}|g||\psi|^4.
\end{eqnarray}
Note that the dissipative term cannot be incorporated into the action.

Taking the variational derivative of Eq.~(\ref{s}), we see that
\begin{equation}
\frac{\delta S}{\delta c_{lm}^*} = \int d^3x \frac{\delta S}{\delta \psi^*}
\frac{\partial \psi^*}{\partial c_{lm}^*}.
\end{equation}
We then use the equations of motion, Eq.~(\ref{vareom}), to arrive at our equations for the variational parameters,
\begin{equation}\label{clm}
 \frac{\delta S}{\delta c_{lm}^*}
=\int d^3x ~(- i \Gamma_3 |\psi|^4\psi)
\frac{\partial \psi^*}{\partial c_{lm}^*}.\end{equation}
Similarly we have two equations associated with $\ell_r$ and $\ell_z$,
\begin{equation}\label{lrz}
 \frac{\delta S}{\delta \ell_{r,z}}= \int d^3x ~( -i \Gamma_3 |\psi|^4\psi)
\frac{\partial \psi^*}{\partial \ell_{r,z}}
+h.c.\end{equation}
We use a computer algebra system to perform the integrals in Eqs.~(\ref{clm})-(\ref{lrz}), to arrive at a coupled set of equations for the variational parameters.  Section \ref{sec:gauss} gives explicit examples in the Gaussian case.

In addition to the time-dependent variational approach, we analyze Eq.~(\ref{gpe1}) by discretizing time and space.  We then use a split-step algorithm to numerically calculate the dynamics of a soliton \cite{javanainen2006}. The results of the discretization approach agree with the results of the variational approach.  The comparison is discussed in section~\ref{compare}.
 
\section{Critical Atom Number}\label{sec:numerical}

\begin{figure}[t]
\def\big{\includegraphics[width=\linewidth]{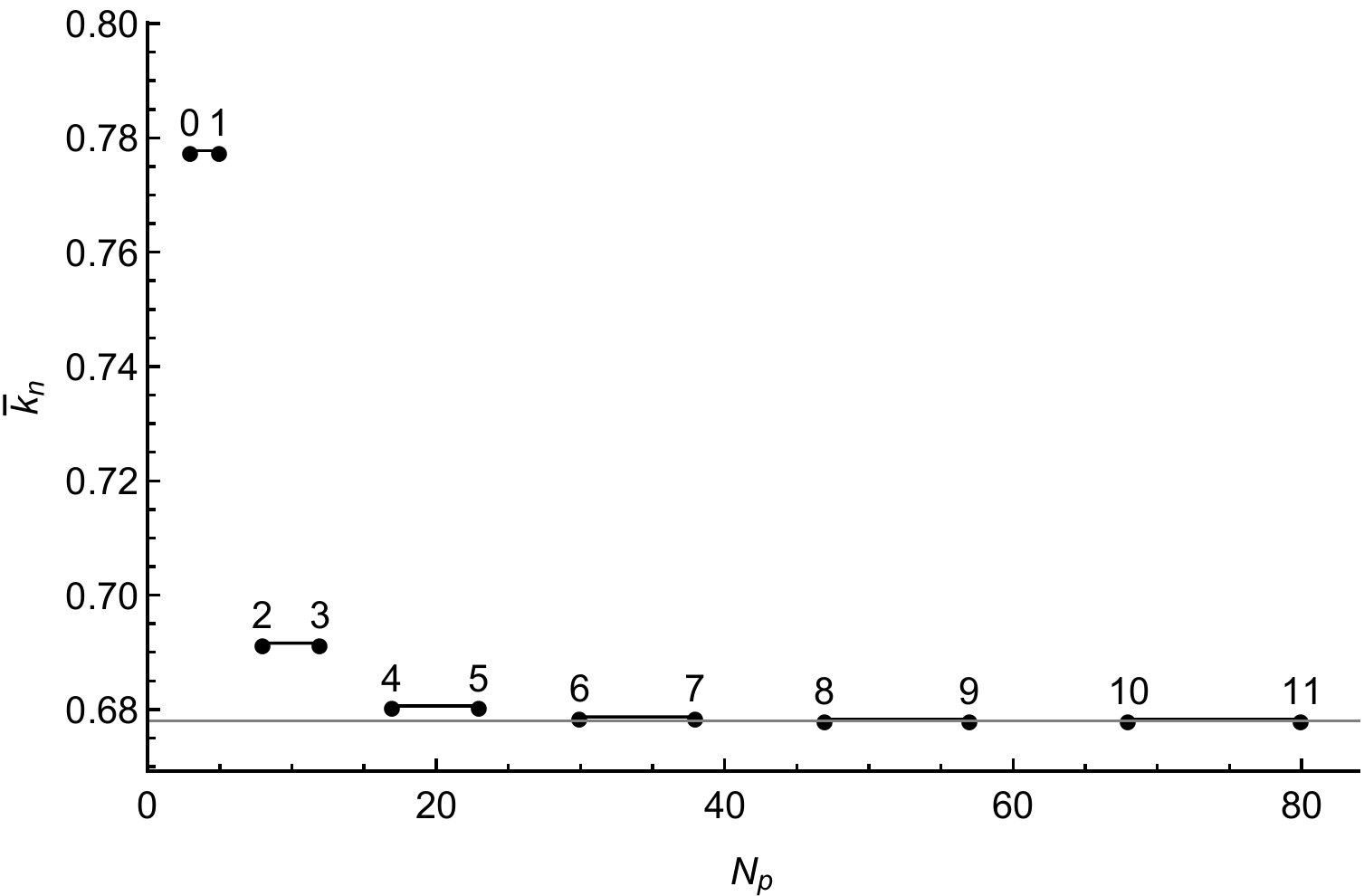}}
\def\little{\includegraphics[height=3.7cm]{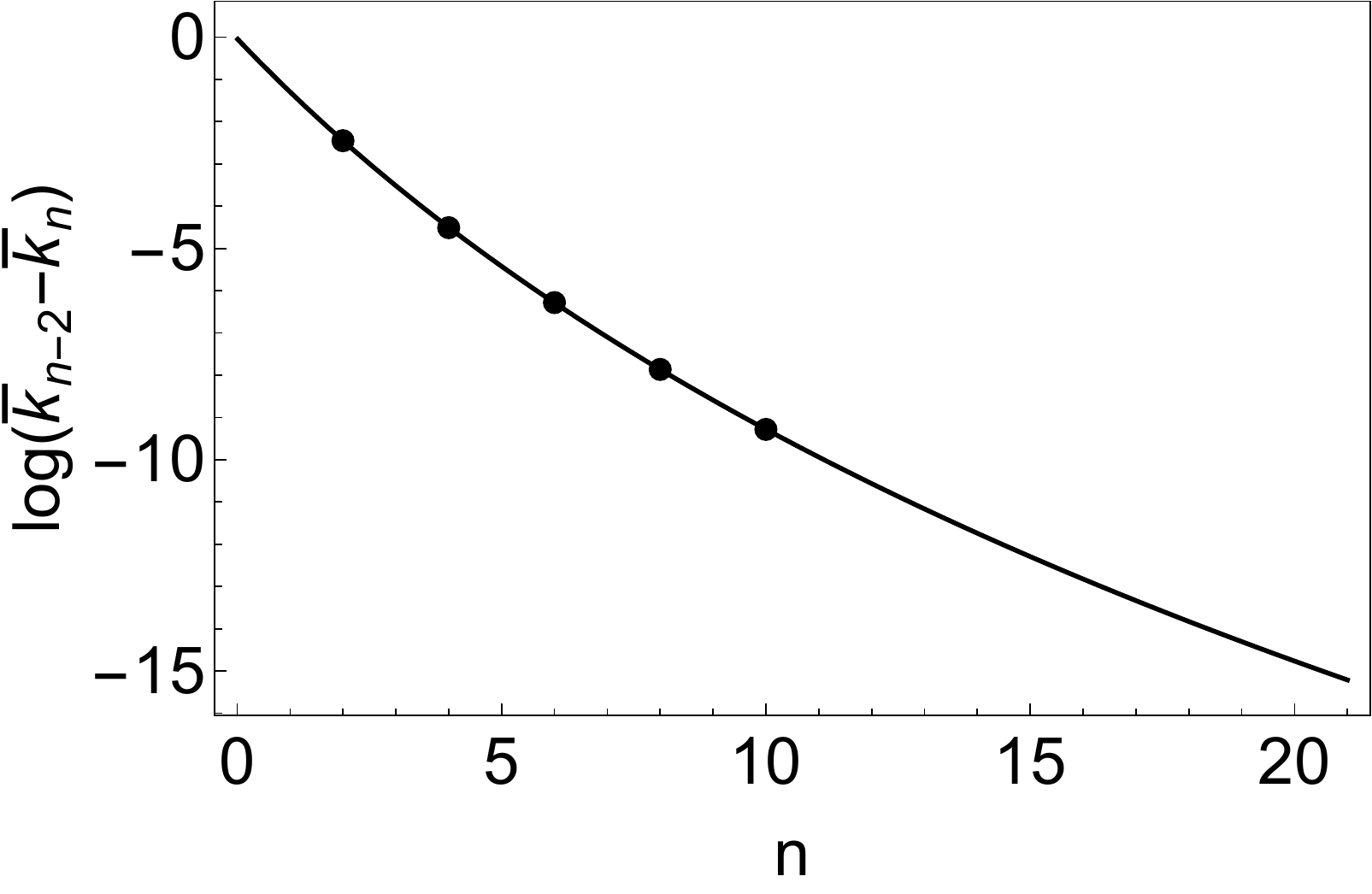}}
\def\stackalignment{r}
\topinset{\little}{\big}{0pt}{0pt}
\caption{\label{criticalnum} Values $\bar k_n$ for different ansatzes are plotted against the corresponding number of parameters, $N_p$. The nearly horizontal black lines convey that $n$ must be increased by 2 to get a better prediction. We do not understand the reason for this pairing behavior.  The horizontal gray line indicates the best fit value of $k_c$. The inset shows the log of the difference between consecutive even-$n$ predictions. The convergence is not quite exponential.}
\end{figure}

As a first application of our variational wavefunction, we calculate $N_c$, the maximum number of atoms which can be found in a static condensate. A soliton with more atoms than $N_c$ is unstable and will collapse. As already explained, the critical atom number can be expressed as $N_c=k_c\sqrt{\frac{\hbar}{m \tilde\omega_r}}\frac{1}{|a_s|}$, where $k_c$ is a dimensionless quantity that only depends on the trap geometry \cite{perezgarcia1998,parker2007}. 
Much theoretical
\cite{perezgarcia1998,parker2007,carr2002,gammal2002,salasnich2002,ruprecht1995} 
and experimental \cite{roberts2001,claussen2003} effort has been devoted to calculating $k_c$. In particular, Parker {\it et al.} \cite{parker2007} calculated that $k_c = 0.675\pm0.005$ for a cylindrical harmonic trap ($\omega_z=0$). They calculated this result with imaginary time evolution of the discretized Gross-Pittaevskii equation, a numerically intensive method. Their result agrees with similar analysis by  Gammal et al. and Salasnich et al.\cite{gammal2002,salasnich2002}.

We calculate the same result using our variational wavefunction, which is a polynomial of degree $2n$ times a Gaussian. Since this is an equilibrium calculation, we can take the coefficients of the polynomials $c_{kj}$ to be real. 
The ansatz will have $N_p =(n+1)(n+2)/2+2$ parameters. We define $\bar k_n$ as the largest value of $k=N |a_s|\sqrt{m\tilde\omega_r/\hbar}$ for which our ansatz yields a soliton solution.  Thus, the best estimate for $k_c$ from a given ansatz is $\bar k_n$.

To find $\bar k_n$, we first choose a modest $k$, which we know is below $\bar k_n$.  We then optimize the parameters of our ansatz by minimizing the energy, using a conjugated gradient method.  Subsequently, we slightly increase $k$  and repeat the minimization, using our previous solution as the starting point.  We continue in this manner until a local minimum cannot be found -- at that point we know we have $k\approx\bar k$.  We further refine our estimate of $\bar k_n$ by using sequentially smaller step sizes.   

The resulting $\bar k_n$ values are shown in Fig.~\ref{criticalnum}.  Remarkably, we find nearly identical results for $\bar k_n$ when  $n=2m$ and $n=2m+1$. 
The even-$n$ data is well approximated by a power law, $\bar k_n = k_c + a(n - b)^{-c}$. We fit the parameters $a, b$, and $c$ in log space, 
by minimizing $\chi^2= \sum_n \log[\bar k_{n-2}-\bar k_n]-\log[a (n-2-b)^{-c}-a (n-b)^{-c}]$.  The fit is illustrated by the inset of Fig.~\ref{criticalnum}. 
We then extract $k_c$, by requiring that our ansatz pass through the final data point.
The uncertainty is estimated by
repeating the fit with one fewer data point, which gives an uncertainty in $k_c$ of $1.6 \times 10^{-6}$. 
We further confirmed the extrapolation by fitting the odd-$n$ data, and by considering a number of alternative functional forms. In all 
cases the results differed by numbers of order $10^{-7}$ or smaller.
 We conclude that $k_c = 0.677986(2)$, corresponding to the horizontal gray line in Fig.~\ref{criticalnum}.

A comparison of the value of $N_c$ for different values of $n$ gives a good estimate for the accuracy of each ansatz. From the data, the Gaussian ansatz has a 15\% error, the n=4 ansatz has a 0.4\% error, and the n=8 ansatz has a 0.02\% error. 

\begin{figure}[t]
\newlength{\awid}
\settowidth{\awid}{(a)}
 (a)\hspace{-\awid}
\raisebox{0.1in-\height}{\includegraphics[width=0.97\linewidth]{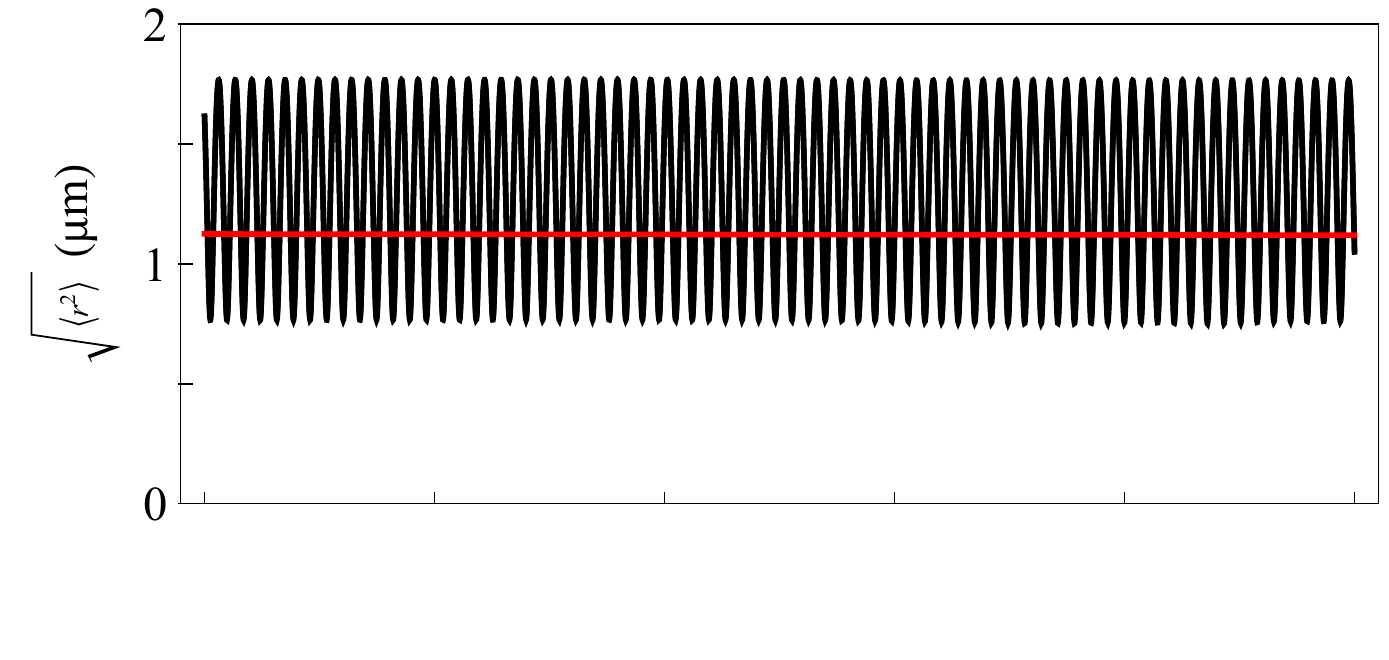}}
\\\
\label{fig:subim1}
\vspace{-0.27in}
\settowidth{\awid}{(b)}
(b)\hspace{-\awid}
\raisebox{0.1in-\height}{\includegraphics[width=0.97\linewidth]{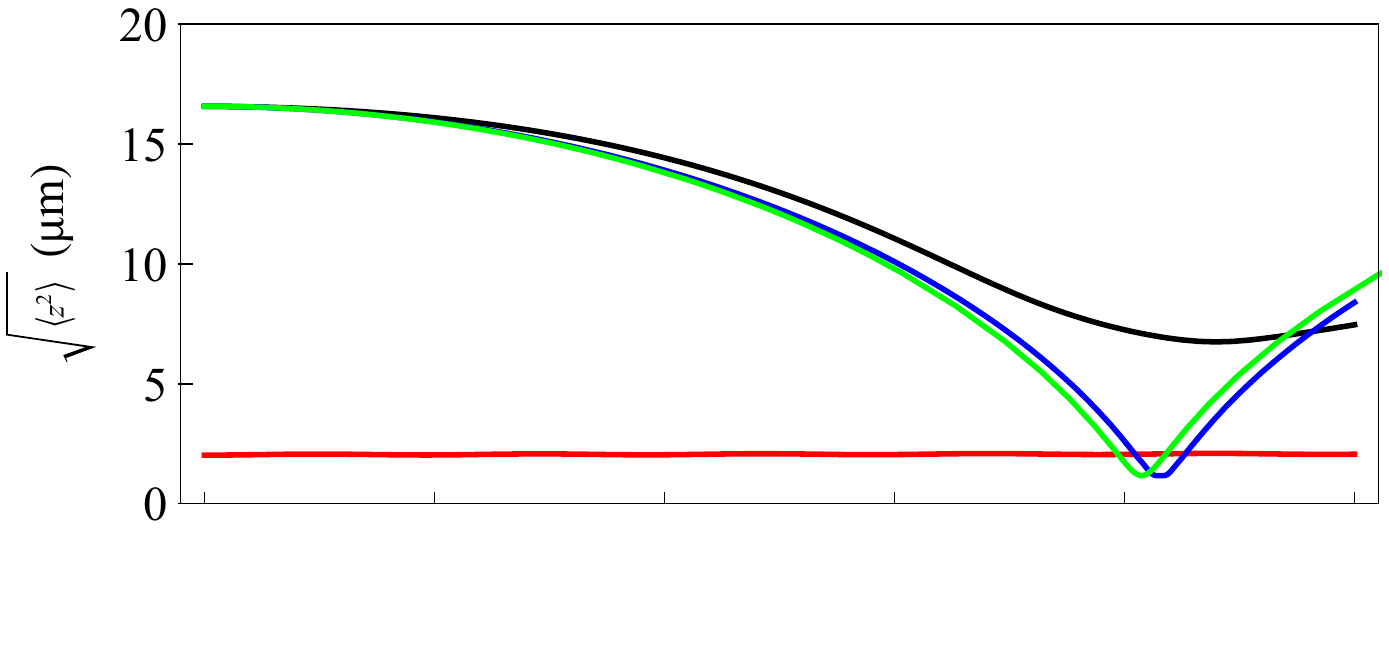}}
\\
\label{fig:subim2}
\vspace{-0.27in}
\settowidth{\awid}{(c)}
(c)\hspace{-\awid}
\raisebox{0.1in-\height}{\includegraphics[width=0.97\linewidth]{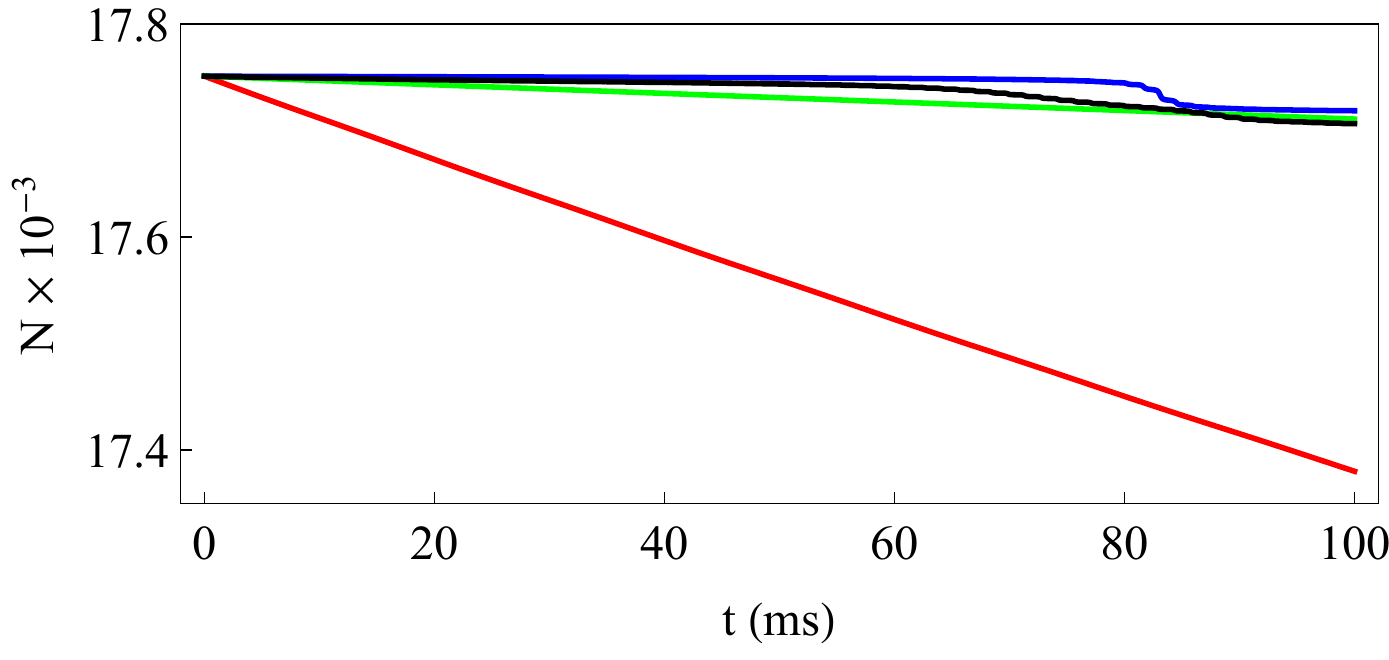}}
\\
\label{fig:subim3}
\vspace{-0.1in}
\caption{\label{osc} Dynamics of $\sqrt{\langle r^2\rangle}$, $\sqrt{\langle z^2\rangle}$ and atom number $N$ for a soliton with highly excited (black, blue, and green) or quasi-stationary (red) initial conditions. The black and red curves were calculated using the variational wavefunction in Eq.~(\ref{ansatz1}), with $n=2$. The blue curves were simulated in the Gaussian approximation ($n=0$), and the green curves show our analytic solutions to the Gaussian ansatz equations, based upon an adiabatic decoupling scheme. In (a), the black, green, and blue curves are all on top of one-another.
Parameters:  $\tilde\omega_r/(2 \pi) =346$ Hz, $\omega_z =0$, $a_s = -0.18 a_0$, which gives $N_c=26130$, and $\Gamma_3 = 6.4 \times 10^{-20}$ cm$^6$/s.}
\label{fig:image2}
\end{figure}

\section{Dynamics}

In Sec.~\ref{sec:gauss} we analyze the equations of motion for the Gaussian ansatz. In particular, in Sec.~\ref{sec:decoupled} through \ref{al} we derive 
 approximate analytic solutions to the equations of motion, and in Sec.~\ref{bd} we describe their limitations.  In section~\ref{compare}, we go beyond the gaussian approximation, calculating dynamics for the ansatz with larger $n$, and compare those results to a numerical integration of the Gross Pittaevskii equation.

\subsection{Gaussian Ansatz}\label{sec:gauss}
The most important parameters describing a soliton are the dimensionless width $\ell_r$, length $\ell_z$, and particle number $|c_{00}|^2=N$.  In this section we analyze the Gaussian ansatz, given by Eq.~(\ref{ansatz1}) with $n=0$. The Gaussian ansatz has six parameters: the above three variables, the global phase of the soliton, $\arg(c_{00}) = \chi$, and the variables $\phi_r$ and $\phi_z$ which account for the probability currents. As already explained, we work in units set by the radial harmonic confinement.  In this case the action can be written as
\begin{equation}\label{action}
\begin{aligned}
S&= -\int dt \left(\frac{N}{2 \ell_r^2}+\frac{N\ell_r^2}{2}+N \ell_r^2(2 \phi_r^2 + \dot{\phi_r}) +\frac{N}{4 \ell_z^2}\right.\\&\quad\quad\quad\quad\quad\quad\left.+\frac{N\omega_z^2\ell_z^2}{4}+\frac{1}{2}N\ell_z^2(2 \phi_z^2 + \dot{\phi_z})+\frac{gN^2}{\ell_r^2\ell_z}+N\dot{\chi}\right)\end{aligned}
\end{equation}

Taking the variational derivatives as in Eqs.~(\ref{clm},\ref{lrz}) then yields

\begin{eqnarray}
\label{phir1}\phi_r-\frac{\dot{\ell_r}}{2\ell_r} &=& \frac{\dot{N}}{6N}\\
\label{phiz1}\phi_z-\frac{\dot{\ell_z}}{2\ell_z}&=& \frac{\dot{N}}{6N}\\
\label{lr1}2\ell_r(\dot{\phi_r}+2\phi_r^2)+\ell_r- \frac{1}{\ell_r^3} - \frac{2 g N}{\ell_r^3 \ell_z}&=&0\\
\label{lz1}2\ell_z(\dot{\phi_z}+2\phi_z^2)- \frac{1}{\ell_z^3} - \frac{2 g N}{\ell_r^2 \ell_z^2}&=&-\omega_z^2\ell_z\\
\label{number2}\dot{N}+\frac{2\Gamma_3 N^3}{3 \sqrt{3} \pi^{3}\ell_r^4 \ell_z^2}&=&0\\
\label{phase1}\dot{\chi}+\frac{1}{\ell_r^2}+\frac{1}{2\ell_z^2} + \frac{7 g N}{2 \ell_r^2 \ell_z} &=&0.
\end{eqnarray}

\subsubsection{Decoupling of timescales}\label{sec:decoupled}
Because of the quasi-1D geometry of the soliton, the timescale of the oscillations in the radial direction is much smaller than that of the oscillations in the axial direction
. Both oscillation timescales are much smaller than the timescale of the atom losses.  The separation of these timescales enables an adiabatic decoupling approach \cite{hinch1991}. 

First, the terms on the right side of Eqs.~(\ref{phir1})-(\ref{phase1}) are insignificant and will be neglected. With this approximation in place, The phases,
$\phi_r,\phi_z$, are related to the lengths $\ell_r,\ell_z$ by $\phi = \dot{\ell}/(2\ell)$, and consequently $2\ell(\dot{\phi}+2\phi^2) = \ddot{\ell}$. This observation allows us to combine Eq.~(\ref{phir1}) and Eq.~(\ref{lr1}) and combine Eq.~(\ref{phiz1}) and Eq.~(\ref{lz1}) into second order differential equations with no first-derivative terms:
\begin{equation}
\label{lr2}
\ddot{\ell_r} + \ell_r - \left(1+\frac{2 g N}{\ell_z}\right)\frac{1}{\ell_r^3}=0.
\end{equation}

\begin{equation}\label{lz2}
\ddot{\ell_z} -\frac{2 g N}{\ell_z^2\ell_r^2}- \frac{1}{\ell_z^3}=0.\end{equation}

Now we can take advantage of the different time-scales in the problem. Since $\ell_z$ and $N$ are slowly varying compared to $\ell_r$, we first solve Eq.~(\ref{lr2}) with $\ell_z$ and $N$ fixed. We then solve Eq.~(\ref{lz2}), replacing $\ell_r^{-2}$ with its time average, $\left\langle\ell_r^{-2}\right\rangle$. As we will soon see, $\left\langle\ell_r^{-2}\right\rangle = 1/\sqrt{1+2 g N/\ell_z} \approx 1$, is not only slowly varying, but also nearly constant.
We will then solve Eq.~(\ref{lz2}) when $g N/\ell_z\ll1$, and substitute those solutions into Eq.~(\ref{number2}) to find $N(t)$.

As we explain in section~\ref{details}, one can map these equations onto two iconic problems in classical mechanics to arrive at analytic solutions.  For $\ell_r$, the solution 
is the distance from the center of an ellipse to its edge,
\begin{eqnarray}\label{lr}
\ell_r^2=&a^2\cos^2{t}+\frac{L^2}{a^2}\sin^2{t}.
\end{eqnarray}
where  $L =\sqrt{1+2g N/\ell_z}$ and $a$ is a parameter which determines the amplitude of motion.  Similarly, $\ell_z(t)$ is the distance from a focus of an ellipse to its edge. The graph is a cycloid, defined parametrically by
\begin{eqnarray}\label{lz}\ell_z(\eta)&=& A(1-\epsilon  \cos{\eta})\\
t(\eta)&=& B(\eta-\epsilon \sin{\eta}).
\end{eqnarray}
Here $\epsilon^2 = 1+{E_z}/({2 g^2 N^2 \left\langle\ell_r^{-2}\right\rangle^2}), A=(2|E_z|(1-\epsilon^2))^{-1/2}, B=(4E_z^2(1-\epsilon^2))^{-1/2}$
and $E_z$ is a parameter determining the amplitude of the motion.

\subsubsection{Solving equations of motion}\label{details}
To find the analytic results in Eqs.~(\ref{lr}) and (\ref{lz}) we note that
Eqs.~(\ref{lz2}) and~(\ref{lr2}) are of the same form as the central force equation:
\begin{equation}\label{classical1}\ddot{r}+\frac{1}{m}\frac{dV(r)}{dr}-\frac{L^2}{m^2r^3}=0\end{equation} 
with $V\propto r^2$ and $V(r)\propto-{1}/{r}$.  These correspond to the harmonic and gravitational potential respectively.
We will find $\vec r(t)$, then our desired solutions are $\ell(t)=r(t) = |\vec r(t)|$.

For the harmonic oscillator, Eq.~(\ref{lr2}), we can characterize the elliptical motion by the semimajor axis of the orbit $a$, and the angular momentum $L =\sqrt{1+2 g N/\ell_z}$.  The energy is $E_r={a^2}/{2}+{L^2}/({2a^2})$. The particle's trajectory can be written as 
\begin{equation}\vec{r}(t) = a\hat{x}\cos{t}+\frac{L}{a}\hat{y}\sin{t}.\end{equation}
Thus we see $\ell_r=|\vec r|$ is given by Eq.~(\ref{lr}).  

For Eq.~(\ref{lz2}), we can characterize the motion of the particle by the energy $E_z$. The equations of motion for $r(t)=\ell_z(t)$ can be integrated with a parametric substitution, as found in Landau's \textit{Mechanics} \cite{landau1982}. The solution is a cycloid:
\begin{eqnarray}\label{lz}\ell_z(\eta)&=& \frac{1-\epsilon \cos{\eta}}{\sqrt{2|E_z|(1-\epsilon^2)}}\\
t(\eta)&=& \frac{\eta-\epsilon \sin{\eta}}{2|E_z|\sqrt{(1-\epsilon^2)}},
\end{eqnarray}
where 
$\epsilon^2 = 1+{E_z}/({2g^2 N^2 \left\langle\ell_r^{-2}\right\rangle^2})$ is the eccentricity of the orbit. 
Bound gravitational orbits, which correspond to self-trapped solitons, have $E_z<0$.
When  $E_z$ approaches zero from below, $\ell_{z,max}$ diverges as $1/|E_z|$ while $\ell_{z,min}$ approaches a constant value. Physically, $E_z > 0$ means that a soliton is no longer self-trapped and can expand to infinite size. However, any nonzero trapping in the axial direction ($\omega_z\neq 0$) would restrain this unlimited expansion.

\subsubsection{Invariants}\label{invar}
One insightful result from the decoupling approximation is the existence of certain invariants, which are found by considering the angular momentum $L=m r^2\dot{\theta}$ of the classical system.  The integral of $\dot{\theta}$ over one period of radial oscillation is $\pi$  or $2\pi$ in the harmonic and graviational problems.
For the radial motion this yields
\begin{equation}\label{aver}
\langle \ell_r^{-2}\rangle = \frac{1}{T_r}
\int_0^{T_r} \frac{dt}{\ell_r^2}=\frac{1}{\sqrt{1+\frac{2gN}{\ell_z}}}\approx 1,
\end{equation}
where we have used that the period of oscillation is $T_r=\pi$.  For the axial motion we instead find
\begin{equation}\label{avez}
T_z\langle \ell_z^{-2}\rangle=
\int_0^{T_z} \frac{d t}{\ell_z^2}=2\pi,
\end{equation}
where $T_z=2\pi/[4 g^2 N^2 (1-\epsilon^2)^{3/2} \langle \ell_r^{-2}\rangle^2]$ is the period of the axial motion. 

We have three applications of these invariants.  First, we used Eq.~(\ref{aver}) in our adiabatic decoupling.  Second,  these invariants are useful in analyzing the evolution of the phase $\chi$. Though $\chi$ is a physically irrelevant global phase for the case of the single soliton, in the context of a soliton train, phase differences between adjacent solitons affect the strength and sign of the soliton-soliton interaction. The phase evolution in Eq.~(\ref{phase1}) is dominated by terms proportional to $\langle \ell_r^{-2}\rangle$, which are independent of the amplitude of radial oscillation.  Consequently the phase relationship between neighboring solitons in a train will be robust against the oscillation of individual solitons.  There is some experimental evidence for this phenomenon \cite{hulet2017}.

Finally, these invariants can be used to analyze the atom loss in Eq.~(\ref{number2}).  The time averaged loss   is proportional to $\langle\ell_z^{-2}\rangle \approx 4 g^2 N^2 (1-\epsilon^2)^{3/2}$. 

\subsubsection{Atom Loss}\label{al}
As explained in the introduction, the experimentally observed atom loss rate is orders-of-magnitude larger than one would expect for static solitons.  We investigated whether or not the oscillations of the solitons were responsible for this discrepency.  Using Eqs.~(\ref{aver}) and (\ref{avez}), the average of $\dot N$ over these oscillations is
\begin{equation}\label{ndot}
\dot N \approx N^5 E_r (1-\epsilon^2)^{\frac{3}{2}} 8\Gamma_3 g^2 /{3 \sqrt{3} \pi^3}
\end{equation}
The atom loss rate depends on the amplitudes of radial and axial oscillation through $E_r$ and $\epsilon$, respectively.  In the absence of oscillations, $\epsilon= 0$ and $E_r=1+{\cal O}(gN/\ell_z)$.  Axial oscillations increase $\epsilon$, and hence decrease $\dot N$.  Radial oscillations increase $E_r$ and hence increase $\dot N$.  This dependence, however, is weak.  If $(\ell_z)_{\rm max}/(\ell_z)_{\rm min}=2$ one has $\epsilon^2=1/9$, while if $(\ell_r)_{\rm max}/(\ell_r)_{\rm min}=2$ one has $E_r=1.25$.  We conclude that the oscillations cannot explain the orders-of-magnitude enhancement of the loss in \cite{hulet2017}.

We can integrate Eq.~(\ref{ndot}) over time-scales large compared to the oscillation period:

\begin{equation}\label{n}
N(t) \approx \frac{N(0)}{\left(1+\frac{4 N(0)^4 (2g)^2 (1-\epsilon_z^2)^{\frac{3}{2}}E_r}{3 \sqrt{3} \pi^3}2\Gamma_3 t\right)^{\frac{1}{4}}}.
\end{equation}

Fig.~\ref{osc}c shows the atom loss in a soliton undergoing a large amplitude axial oscillation with $\epsilon=0.87$. Thus, the dynamic soliton loses atoms at a lower rate: the green curve in the plot, given by Eq.~(\ref{n}), decreases more slowly than the red curve, which represents the atom number of a non-oscillating soliton.  

\subsubsection{Beyond Decoupling}\label{bd}

To study corrections to the decoupling approximation, we numerically integrate Eqs.~(\ref{phir1}) through (\ref{phase1}).  In Fig.~\ref{osc} the blue lines show the numerical solution within the Gaussian approximation, while the green lines show the closed-form predictions from Eqs.~(\ref{lr}), (\ref{lz}), and~(\ref{n}). 

Though these closed-form solutions are fairly accurate for large ranges of the parameters, they do not explain certain features of the dynamics. First, the soliton loses atoms at the greatest rate when the cloud is smallest during each oscillation. This short timescale behavior is not captured in Eq.~(\ref{n}), but when averaged over one period, the closed-form prediction is quite accurate.

Second, the analytic solutions do not capture the slow equilibration between the radial and axial motions. The interaction term can slowly shift energy between the radial and axial modes, until their amplitudes have reached equilibrium. With the current experimental conditions, though, the soliton degrades long before its motion approaches equilibrium. Thus, this effect is not currently experimentally observable.  
 
It is possible to find perturbative analytic solutions which capture these two behaviors. However, for this level of detail, it is more illuminating to investigate the features of the full numerical results.

\subsection{Beyond Gaussian Approximation}\label{compare}

\begin{figure}[t]
\includegraphics[width=\columnwidth]{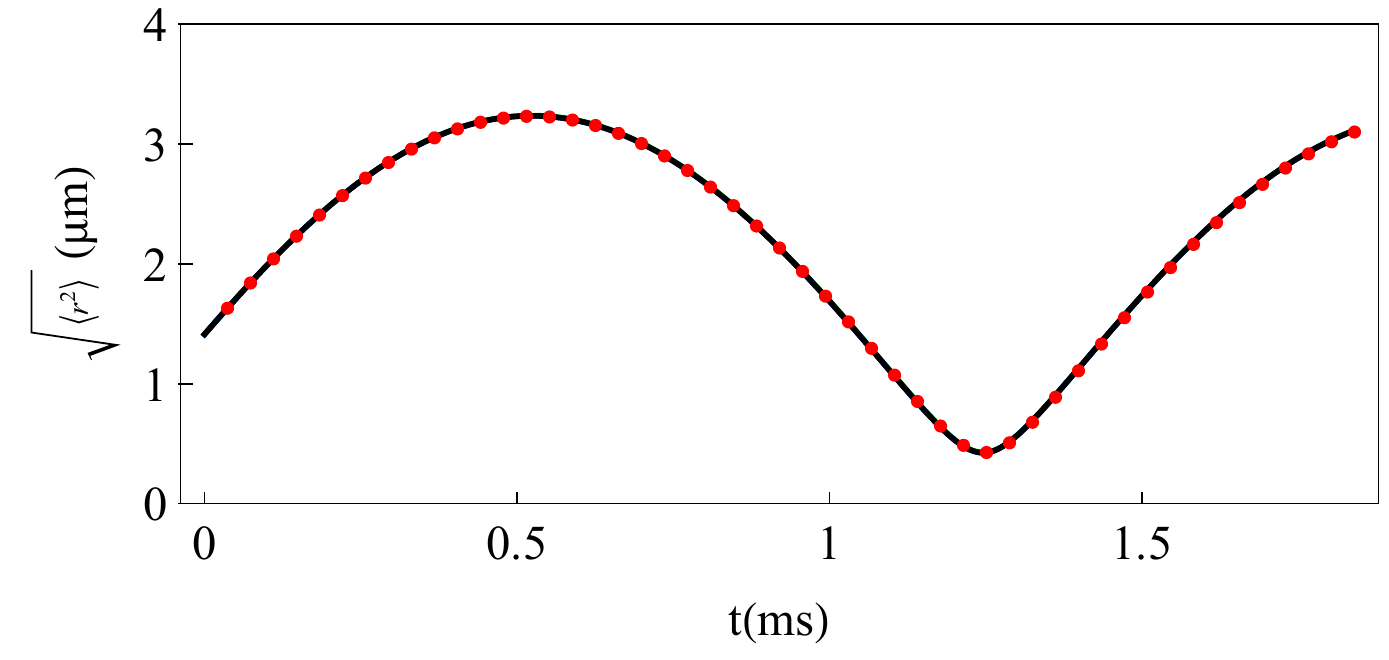}
\caption{\label{discretization} Radius $\sqrt{\langle r^2 \rangle}$ calculated with $n=2$ variational ansatzes (black line), spatially discretized finite difference approach (red dots), the closed-form result~(\ref{lr}) (green) and the Gaussian n=0 ansatz (blue). The green and blue lines are almost entirely hidden behind the n=2 black line, because the errors are small.}
\end{figure}

\begin{figure*}[t]
\label{fig:subim1}
(a)\hspace{-\awid}\hspace{0.08in}\raisebox{0.1in-\height}{\includegraphics[width=0.963\textwidth]{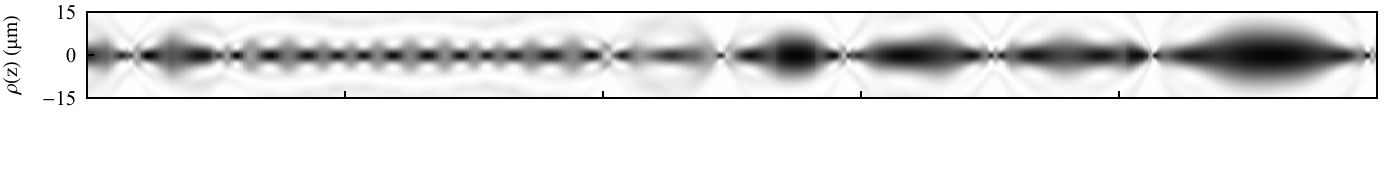}}
\\
\label{fig:subim2}
\vspace{-0.27in}
(b)\hspace{-\awid}\hspace{0.08in}\raisebox{0.1in-\height}{\includegraphics[width=0.963\textwidth]{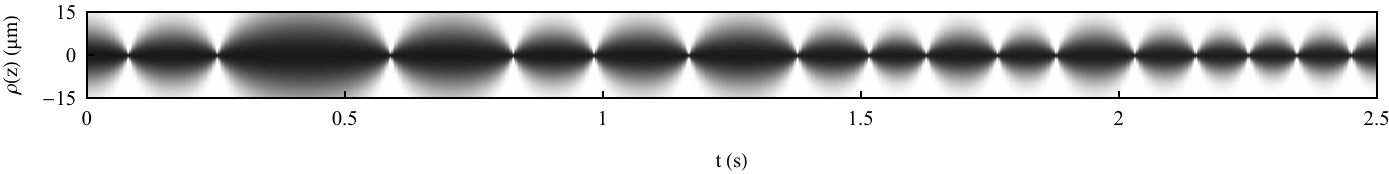}}
\caption{\label{density} The atom density integrated over the radial coordinate,  $\rho(z)=\int\!d^3r n(r,z)$,  plotted over a long timescale for large amplitudes of excitation. Darker colors represent higher axial density. a) The full variational simulation (n=2, 16 real parameters) shows the rich dynamics with multiple timescales. b) Gaussian ansatz. }
\end{figure*}

To more accurately model the dynamics we integrate the coupled ordinary differential equations produced by Eqs.~(\ref{clm}) and (\ref{lrz}) for more complicated ansatzes with larger values of $n$.  Fig.~\ref{osc} shows an example simulation comparing an ansatz with $n=2$ (with 6 complex variational parameters controlling the functional form) to the Gaussian ansatz (with 1 complex variational parameter controlling the functional form) and to the case where the soliton is not oscillating.  

The radial dynamics in panel (a) of Fig.~\ref{osc} show quantitative agreement between the analytic solution and the full simulations. This agreement is also underscored in Fig.~\ref{discretization}, where we compare our variational calculations to a numerical method based on discretizing Eq.~(\ref{gpe1}). This discretization approach follows \cite{javanainen2006} and uses a split-step integrator, taking an adimensional spatial grid spacing of 0.05 and a timestep of 0.001. We ran the simulation over only one radial oscillation because of the computational intensity of the discretization method. Fig.~\ref{discretization} validates the variational results, since they are nearly indistinguishable from the results of spatial discretization.

For the axial dynamics in panel (b) of Fig.~\ref{osc}, quantitative predictive power is lost after a single axial oscillation.  The decoupling approximations break down when the soliton is sufficiently compressed. To illustrate this breakdown, in Fig.~\ref{density}a, we show a longer simulation using the n=2 ansatz (16 real variational parameters) with the same initial conditions as in Fig.~\ref{osc}. We use a density-plot to show the "axial density", $\rho(z)=\int\!d^3r n(r,z)$ as a function of time. For an excitation of this amplitude, the dynamics are quite rich. Energy moves between the various degrees of freedom, and it is likely that the behavior is chaotic: the variations in the amplitude cause large variations in the period of oscillation, which in turn affect the amplitude. Fig.~\ref{density}b shows the results of using the Gaussian ansatz, which does not contain enough degrees of freedom to capture the richness of the dynamics.  The Gaussian approximation is, however, reliable over shorter timescales or for small amplitudes.

\section{Summary}

To summarize, we studied a systematic set of variational ansatzes for the static and dynamic properties of bright solitons in an attractive Bose gas in a cylindrically symmetric trap.  We presented a precise variational calculation of $k_c = N_c|a_s|\sqrt{m \tilde\omega_r/\hbar}$, where $N_c$ is the critical atom number, above which a single soliton will collapse. We found that for a cylindrical harmonic potential, $k_c = 0.677986(2)$.  We studied the dynamics, and found closed-form solutions to the equations from the simplest ansatz by a formal mapping to two familiar classical systems. This simple solution captures all of the significant qualitative features and some of the quantitative features of the far-from-equilibrium dynamics of a bright soliton. 

The oscillations of the soliton affect the atom loss, causing it to occur episodically. When time-averaged, however, we find that the loss rate only depends weakly on the oscillation amplitude.  We conclude that the surprisingly large loss rate seen in the experiments [8] cannot be explained by this dynamical effect.

Although we solely considered the behavior of a single soliton, the generalization to a soliton train is straightforward.  Similar techniques could also be used to explore the soliton generation process, where a monolithic atomic cloud breaks up into an array of solitons.

\section*{Acknowledgements}
We would like to acknowledge Randy Hulet for discussions about the experiments and the physics of cold atom solitons, Junkai Dong for discussions of both techniques and phenomena, and Matthiew Reichl for providing some of the computer code which formed the basis for our spatial discretization approaches.
This material is based upon work supported by the National Science Foundation under Grant No. PHY-1806357 and the ARO-MURI Non-equilibrium Many-body Dynamics Grant No. W9111NF-14-1-0003.

\end{document}